\documentclass[fleqn,10pt]{wlscirep}
\usepackage{amssymb,amsmath,bm}
\usepackage{enumitem}
\usepackage{epsfig,graphics}
\usepackage{color}
\usepackage{graphicx}% Include figure files
\usepackage{dcolumn}% Align table columns on decimal point
\usepackage{bm}% bold math
\usepackage{amsthm}
\usepackage{float}
\usepackage[utf8]{inputenc}
\usepackage[T1]{fontenc}

\renewcommand\thesection{\arabic{section}}

\begin{document}
%\begin{CJK*}{GBK}{song}

\title{Epidemic analysis of COVID-19 in China by dynamical modeling}

\author{Liangrong Peng$^{1*}$, Wuyue Yang$^{2}$\footnote{Those authors contribute equally to this work.}, Dongyan Zhang$^{3}$, Changjing Zhuge$^{3\dag}$, Liu Hong$^{2}$\footnote{Author to whom correspondence should be addressed. Electronic mail: zcamhl@tsinghua.edu.cn(L.Hong), zhuge@bjut.edu.cn(C.Zhuge)}\\
$^1$College of Mathematics and Data Science, Minjiang University, Fuzhou, 350108, P.R.C.\\
$^2$Zhou Pei-Yuan Center for Applied Mathematics, Tsinghua University, Beijing, 100084, P.R.C.\\
$^3$Beijing Institute for Scientific and Engineering Computing, College of Applied
Sciences, Beijing University of Technology, Beijing, 100124, P.R.C.}
%\homepage{$^{*}$Author to whom correspondence should be addressed. Electronic mail: zcamhl@tsinghua.edu.cn(L.H.), zhuge@bjut.edu.cn(C.Zhuge)}% Title
%\footnote{$^*$ The co-authors contribute equally to this work.}

\date{\today} % Date

\begin{abstract}
The outbreak of novel coronavirus-caused pneumonia (COVID-19) in Wuhan has attracted worldwide attention.
Here, we propose a generalized SEIR model by including the self-protection and quarantine to analyze this epidemic. Based on the public data of National Health Commission of China from Jan. 20th to Feb. 16th, 2020, we reliably estimate key epidemic parameters and make predictions on the inflection point and possible ending time of 24 provinces in Mainland and 16 counties in Hubei province. By an optimistic estimation, the epidemics in Beijing and Shanghai will end soon within two weeks (referring to Feb. 16th), while for the most parts of China, including the majority of cities/counties in Hubei province, the success of anti-epidemic will be no later than the middle of March. The situation in Wuhan is still very severe, at least based on the public data until Feb. 16th. We expect it will end up at the beginning of April. Our predictions have been proven to be well in agreement with the real situation. Moreover, by inverse inference, we find the outbreak of COVID-19 in Mainland, Hubei province and Wuhan all can be dated back to the end of December 2019, and the doubling time is around two days in the early stage.
\end{abstract}


%----------------------------------------------------------------------------------------
\maketitle

\section{Introduction}
A novel coronavirus, formerly called 2019-nCoV, or SARS-CoV-2 by ICTV
(severe acute respiratory syndrome coronavirus 2, by the International Committee on Taxonomy of Viruses) caused an outbreak of atypical pneumonia, now officially called COVID-19 by WHO (coronavirus disease 2019, by World Health Organization) first in Wuhan, the capital city of Hubei province in December, 2019 and then rapidly spread out in the whole country of China \cite{Huang2020}.
Till 24:00 Feb. 13th, 2020 (Beijing Time), there are over 60, 000 reported cases (including more than 1, 000 death report) in China, among which, over 80\% are from Hubei province  and over 50\% from Wuhan city \cite{NHC20200213,HUBEI20200213}.

To prevent the spreading of COVID-19, many cities in Hubei province have been locked down since Jan. 23rd, 2020. The central government of China as well as all local governments, has taken a series of very strict preventive measures, such as tracing close contacts, quarantining infected cases, promoting social consensus on self-protection like wearing face mask in public area, \textit{etc.} However, until the finishing of the first version (Feb. 17th, 2020), the epidemic is still ongoing and the daily confirmed cases maintain at a high level.

During this anti-epidemic battle, besides medical and biological research, theoretical studies based on either statistics or mathematical modeling may also play a non-negligible role in understanding the epidemic characteristics of the outbreak, in forecasting the inflection point and ending time, and in deciding the actions to stop the spreading.

For this purpose, in the early stage  many efforts have been devoted to estimate key epidemic parameters, such as the basic reproduction number, doubling time and serial interval, in which the statistical models are mainly used \cite{Muniz-Rodriguez2020,Yang2020,Zhao2020,Sanche2020,Nishiura2020b,Lai2020}.
Due to the limitations of detection methods and restricted diagnostic criteria, asymptomatic or mild patients are possibly excluded from the confirmed cases.
To this end, some methods have been proposed to estimate untraceable contacts \cite{Nishiura2020}, undetected international cases \cite{DeSalazar2020}, or the actual infected cases in Wuhan and Hubei province based on statistical models \cite{Zhao2020c}, or the epidemics outside Hubei province and overseas \cite{Lin2020,Nishiura2020a,Kucharski2020,Zhao2020}.
With the improvement of clinic treatment of patients as well as  the adoption of more strict methods for preventing the spread, many researchers investigate the effect of such changes by statistical reasoning \cite{Chinazzi2020,Jin2020} and stochastic simulations \cite{Hellewell2020,Quilty2020}.

Compared with statistical methods \cite{Zeng2020,huang2020data}, mathematical modeling based on dynamical equations \cite{Kucharski2020,Read2020,Tang2020a, Tang2020} receive relatively less attention, though they can provide more detailed insights into the dynamics of epidemic.
Among them, the classic {\em susceptible exposed infectious recovered} (SEIR) model is the most widely adopted one for characterizing the epidemic of COVID-19 outbreak in both China and other countries \cite{Labadin2020}.
Based on the SEIR model, one can assess the effectiveness of various measures since the outbreak \cite{Shen2020,Tang2020,Tang2020a,Clifford2020,Xiong2020},
%by incorporating the hidden and un-quarantined latent and infectious cases during the transmission process,
which seems to be a difficult task for general statistical methods.
The SEIR model is also utilized to compare the influence of the locking-down of Hubei province on the transmission dynamics in Wuhan and Beijing \cite{Li2020b}.
As the dynamical model can reach interpretable conclusions on the outbreak, a cascade of SEIR models are developed to simulate the processes of transmission from infection source, hosts, reservoir to human \cite{Chen2020}.
There are notable generalizations of the SEIR model for evaluation of the transmission risk and prediction on the patient number, each group of which is divided into two subpopulations, the quarantined and non-quarantined \cite{Tang2020,Tang2020a}.
The extension of the classical SEIR model with delays \cite{Chen2020c,Yue2020} is another routine to simulate the incubation period and the period before recovery.
However, due to the lack of official data and the change of diagnostic caliber during the early stage of the outbreak, most early published models were either too complicated to avoid the overfitting problem, or the parameters were estimated based on limited and less accurate data, resulting in questionable predictions.

In this work, we carefully collect the epidemic data from the authoritative sources: the national, provincial and municipal Health Commissions of China (abbreviated as NHC, see \textit{e.g.} http://www.nhc.gov.cn/). Then we follow the routine of dynamical modeling and focus on the epidemics of COVID-19 inside the whole country, including 24 provinces in Mainland and 16 counties in Hubei province (with confirmed cases more than one hundred). In particular, we pay attention to the data of Mainland excluding Hubei province (denoted as Mainland$^*$), Hubei province excluding Wuhan city (Hubei$^*$) and Wuhan. Such a design aims to minimize the influence of Hubei province and Wuhan city on the data set due to their extremely large infected populations compared to other regions. Without further mention, these conventions will be adopted thorough the whole paper.

By generalizing the classical SEIR model, \textit{e.g.} introducing a new quarantined state and considering the effect of preventive measures, key epidemic parameters for COVID-19, like the latent time, quarantine time and effective reproduction number are determined in a relatively reliable way. The widely interested inflection point, ending time and total infected cases in hot regions are predicted and validated through both direct and indirect evidences. Furthermore, by inverse inference, the starting date of this outbreak are estimated. The analysis of overseas countries are still in progress.

%This paper is organized as follows.
%Section 1 is an Introduction.
%In Section 2, we will describe the model, data set and methods.
%The results will be presented in Section 3.
%This work is concluded and discussed in Section 4.

\section{Model and Methods}

\subsection{Generalized SEIR model}
\begin{figure}[ht]
\[
\includegraphics[scale=0.7]{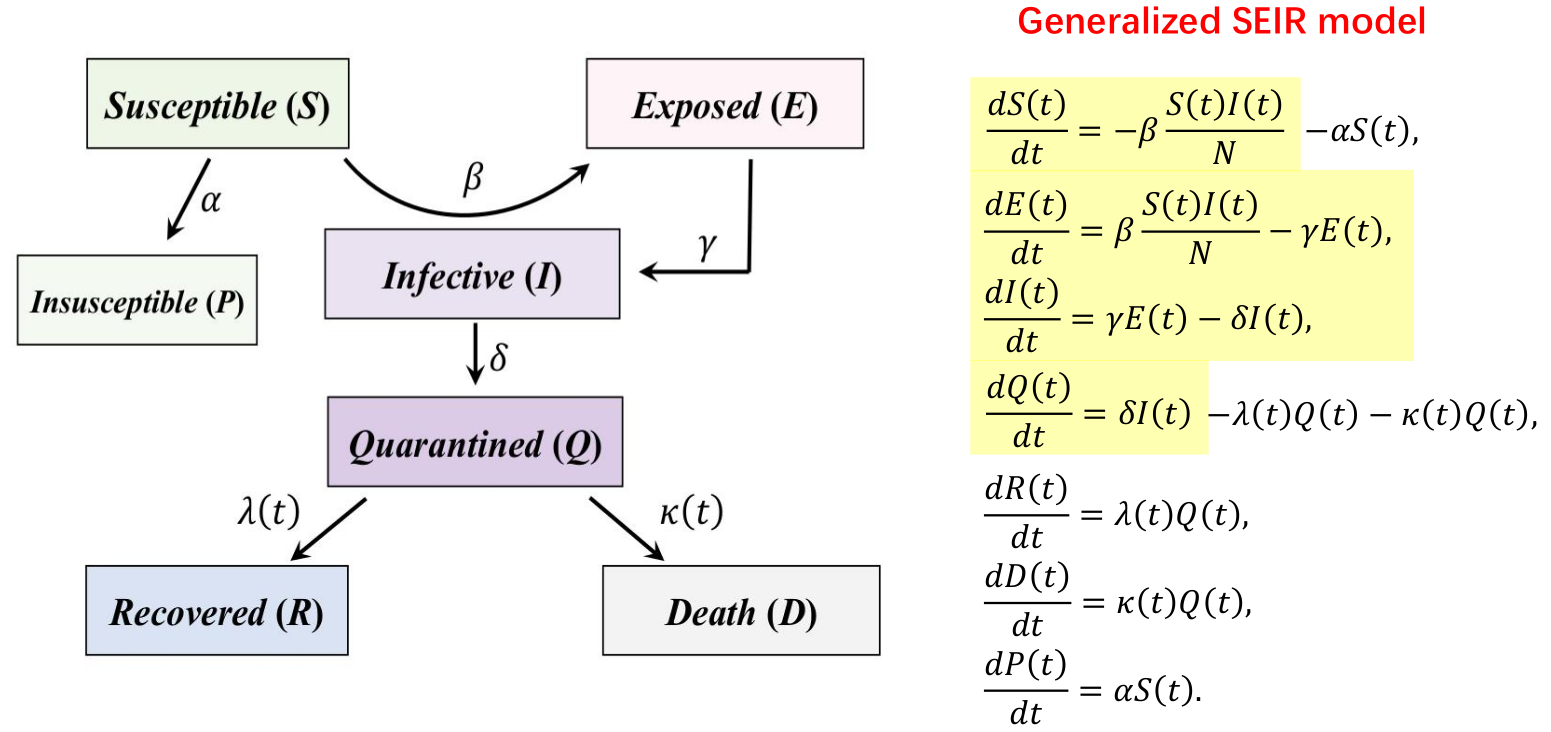}
\]
\caption{The epidemic model for COVID-19. The highlighted part shows the classical SEIR model.}
\label{model}
\end{figure}

%\begin{subequations}\label{m}
%\begin{eqnarray}
%\frac{dS(t)}{dt}&=&-\beta \frac{S(t)I(t)}{N}-\alpha S(t),~\label{m1}\\
%\frac{dE(t)}{dt}&=&\beta \frac{S(t)I(t)}{N}-\gamma E(t), ~\label{m3}\\
%\frac{dI(t)}{dt}&=&\gamma E(t)-\delta I(t), \label{m4}\\
%\frac{dQ(t)}{dt}&=&\delta I(t)-\lambda(t) Q(t)-\kappa(t) Q(t), \label{m5}\\
%\frac{dR(t)}{dt}&=&\lambda(t) Q(t), \label{m6}\\
%\frac{dD(t)}{dt}&=&\kappa(t) Q(t), \label{m7}\\
%\frac{dP(t)}{dt}&=&\alpha S(t).
%\end{eqnarray}
%\end{subequations}
To characterize the epidemic of COVID-19 which outbroke in Wuhan at the end of 2019, we generalize the classical SEIR model\cite{Labadin2020,Shen2020,Tang2020,Tang2020a,Clifford2020,Xiong2020,Li2020b} by introducing seven different states, \textit{i.e.} $\{S(t), P(t), E(t), I(t), Q(t), R(t), D(t)\}$ denoting at time $t$ the respective number of the {\em susceptible cases}, {\em insusceptible cases}, {\em exposed cases} (infected but not yet be infectious, in a latent period), {\em infectious cases} (with infectious capacity and not yet be quarantined), {\em quarantined cases} (confirmed and infected), {\em recovered cases} and {\em closed cases} (or death).
The adding of a new quarantined sate is driven by data, which together with the recovery state takes replace of the original $R$ state in the classical SEIR model. Their relations are given in Fig. \ref{model} and characterized by a group of ordinary differential equations (or difference equations if we consider discrete time, see SI). Constant $N=S+P+E+I+Q+R+D$ is the total population in a closed region. The coefficients $\{\alpha, \beta, \gamma^{-1}, \delta^{-1}, \lambda(t), \kappa(t)\}$ represent the protection rate, infection rate, average latent time, average quarantine time, cure rate, and mortality rate, separately. Especially, to take the improvement of public health into account, such as encouraging wearing face masks, more effective contact tracing and more strict locking-down of communities, we assume that the susceptible population is stably decreasing and thus introduce a positive protection rate $\alpha$ into the model. In this case, the effective reproduction number becomes $ERN=\beta*\delta^{-1}*(1-\alpha)^T$, $T$ is the number of days since the awareness of epidemic.

\begin{figure}[ht]
\[
\includegraphics[scale=0.6]{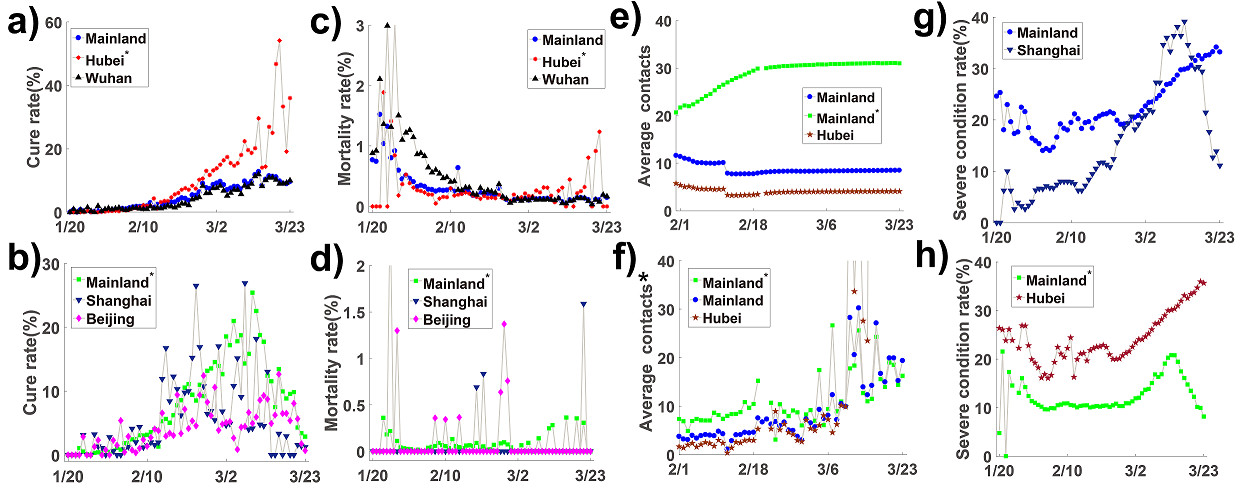}
\]
\caption{(Color online) (a)-(b) The cure rate $\lambda$, (c)-(d) mortality rate $\kappa$, (e)-(f) average close contacts, and (g)-(h) severe condition rate (see SI for their definitions) are calculated based on the public data from NHC of China from Jan. 20th to Mar. 23rd for Mainland, Mainland$^*$, Hubei, Hubei$^*$, Wuhan, Beijing and Shanghai separately.}
\label{parameters}
\end{figure}

It is noted that here we assume the cure rate $\lambda$ and the mortality rate $\kappa$ are both time dependent. As confirmed in Fig. \ref{parameters}a-d, the cure rate $\lambda(t)$ is gradually increasing with the time, while the mortality rate $\kappa(t)$ quickly decreases to less than $1\%$ and becomes stabilized after Jan. 30th. This phenomenon is likely raised by the assistance of other emergency medical teams, the application of new drugs, \textit{etc.} Furthermore, the average contact number of an infectious person is calculated in Fig. \ref{parameters}e-f and could provide some clues on the infection rate. It is clearly seen that the average contact number is basically stable over time after Feb. 10th, but shows a remarkable difference between inside and outside Hubei province, which could be attributed to different quarantine policies and implements, since a less severe region is more likely to inquiry the close contacts of a confirmed case. A similar regional difference is observed for the severe condition rate too. In Fig. \ref{parameters}g-h, Hubei province (including Wuhan) shows a much higher severe condition rate than other parts of China. Although it is generally expected that the patients need a period of time to become infectious, to be quarantined, or to be recovered from illness, we do not find a strong evidence for the necessity of including time delays (see SI for more details). As a result, the time-delayed equations are not considered in the current work for simplicity.

\subsection{Parameter estimation}

According to the daily official reports of NHC of China, the cumulative numbers of quarantined cases, recovered cases and closed cases are available in public. However, since the latter two are directly related to the first one through the time dependent recovery rate and mortality rate, the numbers of quarantined cases $Q(t)$ plays a key role in our modeling. Furthermore, as the accurate numbers of exposed cases and infectious cases are very hard to determine, they will be treated as hidden variables during the study.

Leaving alone the time dependent parameters $\lambda(t)$ and $\kappa(t)$, there are four unknown coefficients $\{\alpha, \beta, \gamma^{-1}, \delta^{-1}\}$ and
two initial conditions $\{E_0, I_0\}$ about the hidden variables (other initial conditions are known from the data) have to be extracted from the time series data $\{Q(t)\}$. Such an optimization problem could be solved automatically by using the least-square regression combined with the simulated annealing algorithm (see SI for details). A major difficulty is how to overcome the overfitting problem.

To this end, we firstly prefix the latent time $\gamma^{-1}$ in accordance with previous reports of 3-7 days \cite{Guan2020,Yang2020,Li2020}. And then for each fixed $\gamma^{-1}$, we explore its influence on other parameters ($\beta=1$ nearly unchanged), initial values, as well as the population dynamics of quarantined cases and infected cases during best fitting. From Fig. \ref{sensitivity}a-b, to produce the same outcome, the protection rate $\alpha$ and the reciprocal of the quarantine time $\delta^{-1}$ (with an average of 4.95 days) are both decreasing with the latent time $\gamma^{-1}$, which is consistent with the fact that longer latent time requires longer quarantine time. Meanwhile, the initial values of exposed cases and infectious cases are increasing with the latent time. Since $E_0$ and $I_0$ include asymptomatic patients, they both should be larger than the number of quarantined cases. Furthermore, as the time period between the starting date of our simulation (Jan. 20th) and the initial outbreak of COVID-19 (generally believed to be earlier than Jan. 1st) is much longer than the latent time (3-7 days), $E_0$ and $I_0$ have to be close to each other, which makes only their sum $E_0+I_0$ matters during the fitting. An additional important finding is that in all cases $\beta$ is always very close to 1, which agrees with the observation that COVID-19 has an extremely strong infectious ability. Nearly every unprotected person will be infected after a direct contact with the COVID-19 patients \cite{Guan2020,Yang2020,Li2020}. As a summary, we conclude that once the latent time $\gamma^{-1}$ is fixed, the fitting accuracy on the time series data $\{Q(t)\}$ basically depends on the values of $\alpha$, $\delta^{-1}$ and $E_0+I_0$.

\subsection{Sensitivity analysis}

\begin{figure}[ht]
\[
\includegraphics[scale=0.4]{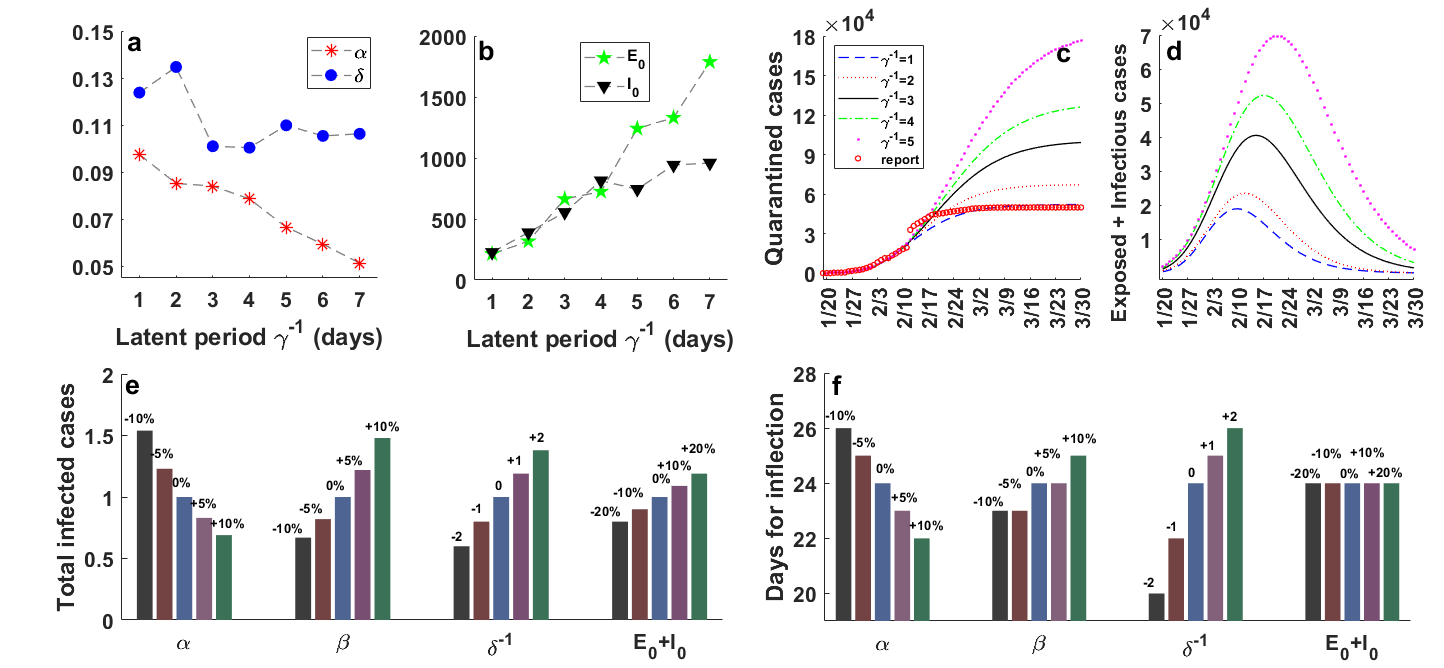}
\]
\caption{(Color online) Sensitivity analysis on parameters for the generalized SEIR model.
The influence of the latent time on
(a) the protection rate $\alpha$ and quarantine time $\delta^{-1}$,
(b) the initial values of exposed cases $E_0$ and infected cases $I_0$ on Jan. 20th,
(c) the cumulative quarantined cases,
(d) the sum of exposed and infectious cases $E(t)+I(t)$, $i.e.$, the currently infected but not yet quarantined cases.
(e) Effects of other parameters on the final total infected cases;
(f) and the time period from the starting point (Jan. 20th) to the inflection point (when the effective reproduction number becomes less than 1).
In the top panels, the value of latent time $\gamma^{-1}$ is varied; while in the bottom panels, $\gamma^{-1}$ is fixed. All calculations are performed with respect to the data of Wuhan city, with reported data (red circles) obtained from NHC of China.
}
\label{sensitivity}
\end{figure}

In order to further evaluate the influence of other fitting parameters on the long-term forecast, we perform sensitivity analysis on the data of Wuhan (results for other regions are similar and not shown) by systematically varying the values of unknown coefficients \cite{huang2015partial, hong2017statistical}. As shown in Fig. \ref{sensitivity}e-f, the predicted total infected cases at the end of epidemic, as well as the the inflection point, at which the effective reproduction number is less than 1\cite{Zhao2020}, both show a positive correlation with the infection rate $\beta$ and the quarantined time $\delta^{-1}$ and a negative correlation with the protection rate $\alpha$. These facts agree with the common sense and highlight the necessity of self-protection (increase $\alpha$ and decrease $\beta$), timely disinfection (increase $\alpha$ and decrease $\beta$), early quarantine (decrease $\delta^{-1}$), \textit{etc.} An exception is found for the initial total infected cases. Although a larger value of $E_0+I_0$ could substantially increase the final total infected cases, it shows no impact on the inflection point, which could be learnt from the previous formula of effective reproduction number.

\section{Results and Discussion}

\subsection{Interpretation of the public data}
We apply our pre-described generalized SEIR model to interpret the public data on the cumulative numbers of quarantined cases from Jan. 20th to Feb. 16th, which are published daily by NHC of China since Jan. 20th. Our study includes 24 provinces in Mainland and 16 counties in Hubei province, whose numbers of reported confirmed cases are more than 100 at that moment.

Through extensive simulations, the optimal values for unknown model parameters and initial conditions, which best explain the observed cumulative numbers of quarantined cases (see Figs. \ref{fig:full-prediction-mainland} and \ref{fig:full-prediction-hubei}), are determined and summarized in tables in SI. There are several remarkable facts could be immediately learnt from those tables. Firstly, the protection rate of Wuhan is significantly lower than other regions, showing many infected cases may not yet be well quarantined until Feb. 16th (the smaller $\alpha$ for Wuhan does not necessarily mean people in Wuhan pay less attention to self-protection, but more likely due to the higher mixing ratio of susceptible cases with infectious cases). Similar lower protection rates are observed in several other counties in Hubei province too. Secondly, a positive correlation between the quarantine time and the level of economic development is observed. The quarantine time for Beijing, Tianjin, Shandong and Guangdong provinces are about 4 days, which significantly increase to more than 7 days for Heilongjiang, Sichuan and Yunnan provinces. Again, the quarantine time for Wuhan is among the longest (around 6.3 days). The short quarantine time for Hebei, Hunan, Guangxi and Hainan could be attributed to extremely strict measures taken in these places. Finally, the estimated number of total infected cases on Jan. 20th in all places are significantly larger than one, suggesting the COVID-19 has already spread out nationwide at that moment. We will come back to this point in the next part.

%\begin{table}[ht]
%\[
%\includegraphics[scale=0.9]{Table.png}
%\]
%\caption{Summary of all constant parameters for the generalized SEIR %model. $E_0$ and $I_0$ denote the initial values for exposed cases and %infectious cases separately. The time-dependent cure rate $\lambda(t)$ %and mortality rate $\kappa(t)$ can be read out from Fig. \ref{parameters} %and are given in SI.}
%\label{table1}
%\end{table}

\subsection{Forecast for the epidemic of COVID-19}

\begin{figure}[tphb]
    \centering
    \includegraphics[width=15cm]{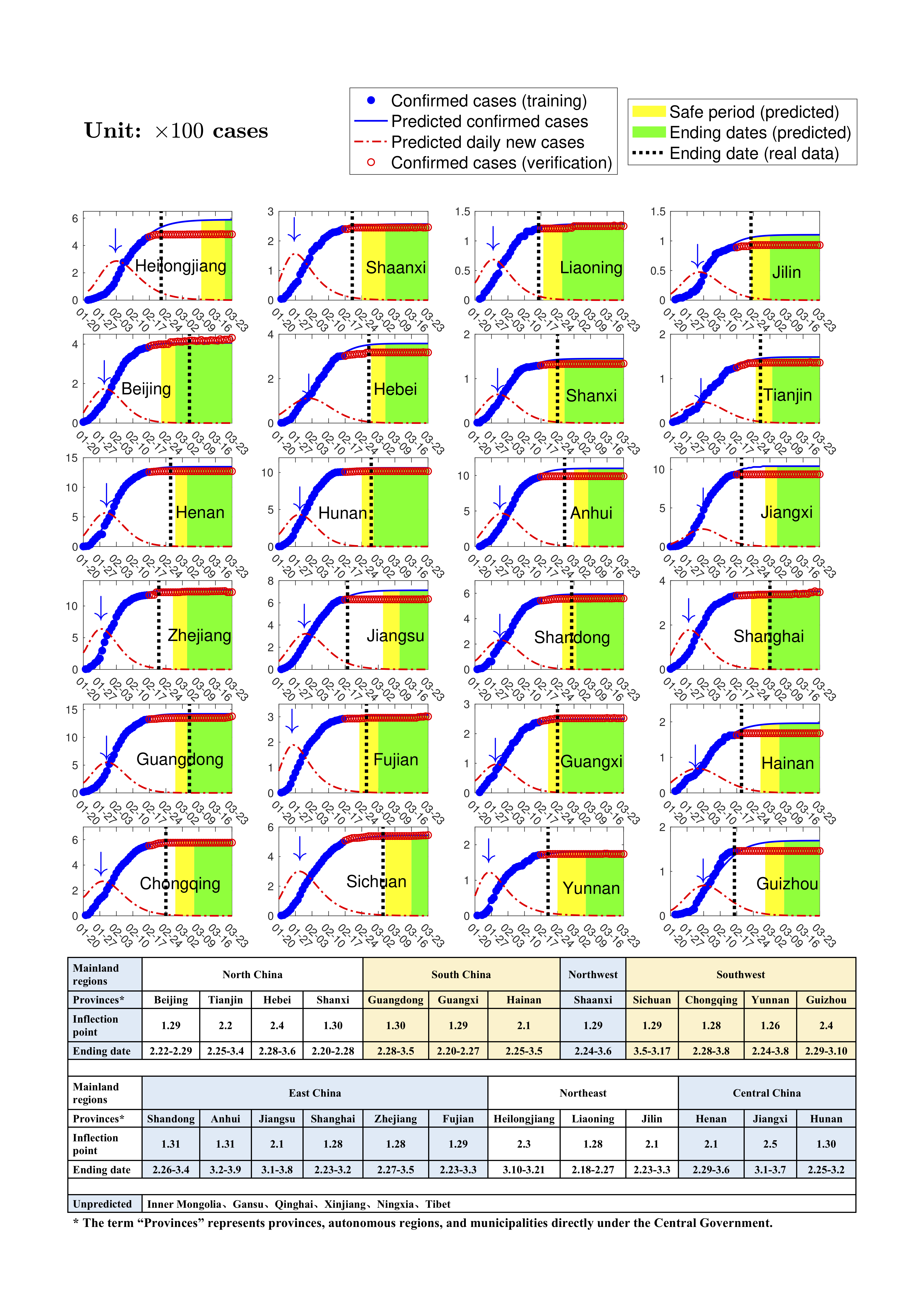}
    \caption{Predictions on the accumulative confirmed cases (the oversea imported cases are not included) of 24 provinces, autonomous regions and/or municipalities in Mainland of China from Feb. 17th to Mar. 23rd. The predicted safe period (yellow zone) is defined as the date, after which there are less than 5 newly confirmed cases, while that of the predicted ending date (green zone) is defined as less than 1 newly confirmed case. The real ending date is defined as the beginning date, from which there are no more new cases for 7 successive days. The table in the bottom summarizes the predicted inflection points and ending dates characterizing the epidemic dynamics.}
    \label{fig:full-prediction-mainland}
\end{figure}

\begin{figure}[tphb]
    \centering
    \includegraphics[width=15cm]{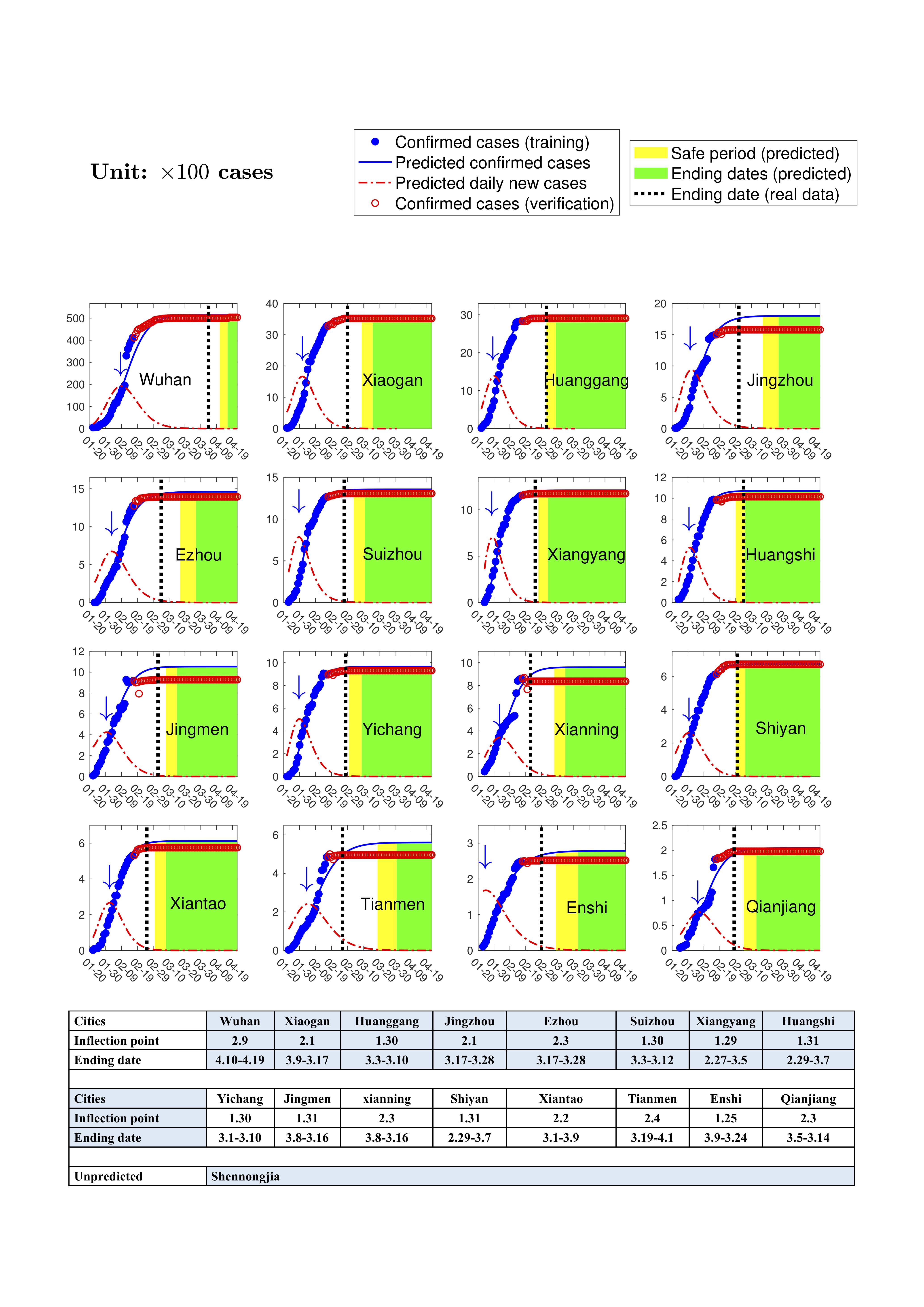}
    \caption{Predictions on the counties in Hubei Province from Feb. 17th to Apr. 19th. Similar to Figure \ref{fig:full-prediction-mainland}. }
    \label{fig:full-prediction-hubei}
\end{figure}

\begin{figure}[ht]
\[
\includegraphics[scale=0.55]{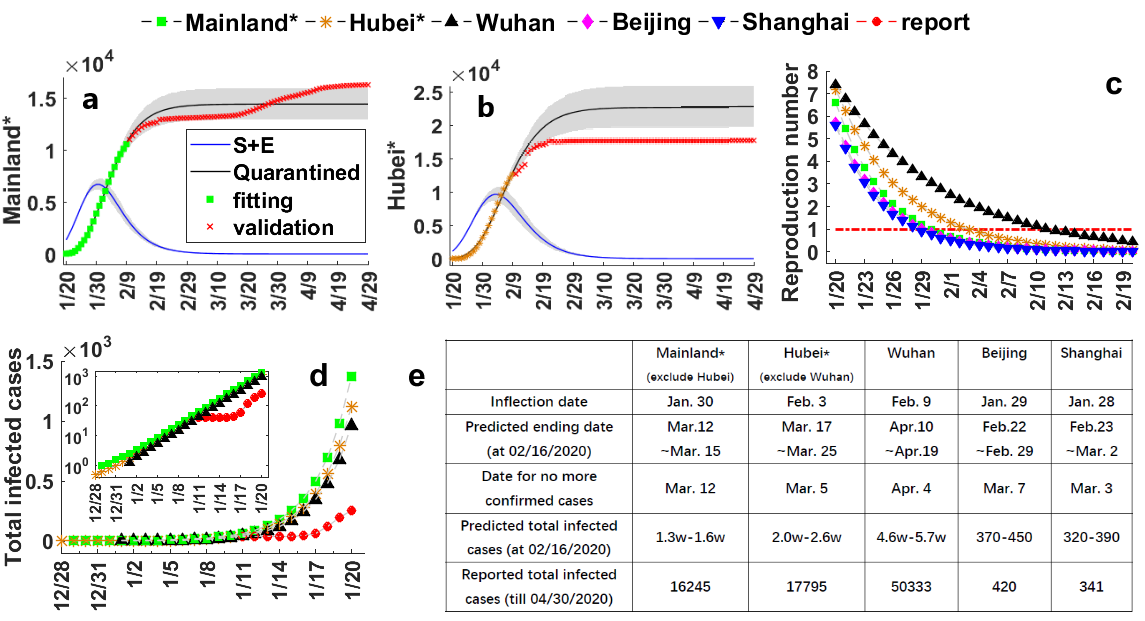}
\]
\caption{
(Color online) (a-b) Predicted cumulative quarantined cases in the mainland (excluded Hubei) and Hubei province (excluded Wuhan) from Feb. 10th to Apr. 30th, 2020. The shaded area indicates a 95\% confidence interval. The data points from Jan. 20th to Feb. 9th, 2020 are taken as fitting, while those from Feb. 10th to Apr. 30th are taken as validation. Parameters are set in accordance with Table 1. (c) The effective reproduction number, (d) the estimated total infected cases at the early stage of COVID-19 epidemic between Dec. 28th, 2019 and Jan. 20th, 2020 by inverse inference, and (e) a summary on the estimated inflection point, ending date and total number of final infected cases in Mainland$^*$, Hubei$^*$, Wuhan, Beijing and Shanghai, in comparison with the official reports.
}
\label{prediction}
\end{figure}

Most importantly, with the model and parameters in hand, we can carry out simulations for a longer time and forecast the potential tendency of the COVID-19 epidemic. In Figs. \ref{fig:full-prediction-mainland} and \ref{fig:full-prediction-hubei}, the cumulative number of quarantined cases in 24 provinces in Mainland and 16 counties in Hubei province are predicted and compared with official data from Feb. 17th to Mar. 23rd (as marked in red spots and taken as a direct validation). Meanwhile, the predicted cumulative number of quarantined cases and the current number of exposed cases plus infectious cases for the Mainland$^*$ and Hubei$^*$ are summarized in Figs. \ref{prediction}a-b.

Overall, our forecasts show a well agreement with the validation data. And we are delighted to see the reported confirmed cases of several provinces, like Heilongjiang, Anhui, Jiangxi, Jiangsu, Hainan, Guizhou, \textit{etc.}, are lower than our predictions. The real ending dates for these provinces are also earlier. Especially, the reported number of confirmed cases in Hubei$^*$ is about $20\%$ lower than our predictions (see Fig. \ref{prediction}b), showing the nationwide anti-epidemic measures come into play. While for Wuhan city (and also Hubei province), due to the inclusion of suspected cases with clinical diagnosis into confirmed cases (12364 cases for Wuhan and 968 cases for Hubei$^*$ on Feb. 12th) announced by NHC of China since Feb. 12th during the preparation of our first version, there is a sudden jump in the quarantined cases. Although it to some extent offsets our original overestimates, it also reveals the current severe situation in Wuhan city, which requires much closer attention in the future.

%\begin{figure}[H]
%\[
%\includegraphics[scale=0.5]{Fig4.png}
%\]
%\caption{(Color online) Predictions of the generalized SEIR model on the cumulative quarantined cases (red solid lines), sum of current exposed and infectious cases (blue solid lines), cumulative recovered cases (purple solid lines), and cumulative closed cases (green solid lines) in Mainland$^*$, Hubei$^*$, Beijing, Shanghai, and Wuhan (from top to bottom). The red triangles,  purple asterisks and green circle represent the public data points between Jan. 20th and Feb. 9th, 2020. The shaded area indicates predictions within 95\% confidence interval. With the Euclidean distance $\parallel \cdot \parallel_2$, the average relative error $RE=\sqrt{\frac{\parallel y-x \parallel_2}{\parallel x\parallel_2}}$ between the prediction $y$ and public data $x$ is evaluated for the cumulative quarantined cases, that is $RE=2.4\%, 5.6\%, 1.9\%, 2.9\%, 3.8\%$ for Mainland$^*$, Hubei$^*$, Beijing, Shanghai and Wuhan.
%Parameters are taken in accordance with Table 1.
%}
%\label{fitting}
%\end{figure}

Towards the epidemics of COVID-19 in China, our basic predictions are summarized as follows:
\begin{itemize}
\item[1.] By an optimistic estimation, the epidemics of COVID-19 in Beijing and Shanghai would soon be ended within two weeks (from Feb. 16th). While for most parts of Mainland, the success of anti-epidemic will be no later than the middle of March. The situation in Wuhan is still very severe, at least based on public data until Feb. 16th. We expect it will end up at the beginning of April. Now after three months since the first version was finished, we are happy to see that our forecasts have been proven to be well in agreement with the real situation.
\item[2.] The estimated final total infected cases (not only total quarantined cases) for Beijing and Shanghai will be around four hundred. This number is about 13-16 thousand for Mainland$^*$ (exclude Hubei province), 20-26 thousand for Hubei$^*$ province (exclude Wuhan city) and 46-57 thousand for Wuhan city. The reported total infected cases till Apr. 30th, 2020 for Mainland$^*$ and Wuhan fall into our predicted regions, while that for Hubei$^*$ is about $20\%$ lower than our estimation due to the effective measures taken by the local governments of Hubei province and the powerful assistance of emergency medical teams from other provinces of China.
\item[3.] According to the effective reproduction number shown in Fig. \ref{prediction}c, the inflection date for Beijing, Shanghai, Mainland$^*$ (exclude Hubei) is around Jan. 30th, which is close to the reported Feb. 3rd for the last one based on daily new confirmed cases. The inflection point for Hubei$^*$ province (exclude Wuhan city) agrees with the reported Feb. 5th. These facts indicate that the epidemic is now under well control in most cities in China.
\item[4.] The predicted inflection point, ending date and final number of total infected cases are summarized in Fig. \ref{prediction}e. In particular, the inflection point for Wuhan city is determined as Feb. 12th (data after Feb. 9th are not included into parameter estimation). By coincidence, on the same day, we witnessed a sudden jump in the number of confirmed cases due to a relaxed diagnosis caliber, meaning more suspected cases will receive better medical care and have much lower chances to spread virus. Besides, Wuhan local government announced the completion of community survey on all confirmed cases, suspected cases and close contacts in the whole city.
\end{itemize}

\subsection{Inverse inference on the outbreak of COVID-19}

Besides the forecast, the early trajectory of the COVID-19 outbreak is also crucial for our understanding as well as for future prevention. To this end, by adopting the shooting method, we carry out inverse inference to explore the early epidemic dynamics of COVID-19 since its onset in Mainland$^*$, Hubei$^*$, and Wuhan (other regions are not considered due to their too small numbers of confirmed cases on Jan. 20th). With respect to the parameters and initial conditions listed in tables in SI, we make an astonishing finding that, for all three cases, the outbreaks of COVID-19 all point to 20-25 days before Jan. 20th (the starting date for public data and our modeling). It means the epidemics of COVID-19 in these regions are no later than Jan. 1st (see Fig. \ref{prediction}d), in agreement with reports by Li \textit{et al. }\cite{Guan2020,Yang2020,Li2020}. And in this stage (from Jan. 1st to Jan. 20th), the number of total infected cases follows a nice exponential curve with the doubling time around 2 days. This in some way explains why statistical studies with either exponential functions or logistic models could work very well on early limited data points. Furthermore, we notice the number of infected cases based on inverse inference is much larger than the reported confirmed cases in Wuhan city before Jan. 20th.

\section{Conclusion}

In this study, we propose a generalized SEIR model to analyze the epidemic of COVID-19, which was first reported in Wuhan last December and then quickly spread out nationwide in China. Our model properly incorporates the intrinsic impacts of hidden exposed and infectious cases on the entire procedure of epidemic, which is difficult for traditional statistical analysis. A new quarantined state, together with the recovered state, takes replace of the original $R$ state in the classical SEIR model and correctly accounts for the daily reported confirmed infected cases and recovered cases.

Based on detailed analysis of the public data of NHC of China from Jan. 20th to Feb. 16th, we estimate several key parameters for COVID-19, like the latent time, the quarantine time and the effective reproduction number in a relatively reliable way, and predict the inflection point, possible ending time and final total infected cases for 24 provinces in Mainland and 16 counties in Hubei province. Overall, the epidemic situations for Beijing and Shanghai are optimistic, which are expected to end up within two weeks (from Feb. 16th, 2020). Meanwhile, for most parts of Mainland including the majority of counties in Hubei province, it will be no later than the middle of March. We should also point out that the situation in Wuhan city is still very severe, at least based on the public data until Feb. 16th. We expect it will end up at the beginning of April. Our predictions have been proven to be well in agreement with the real situation.

Furthermore, by inverse inference, we find that the outbreak of COVID-19 epidemics in Mainland$^*$, Hubei$^*$, and Wuhan can all be dated back to 20-25 days ago with respect to Jan. 20th, in other words the end of Dec. 2019, which is consistent with public reports. Although we lack the knowledge of the first infected case, our inverse inference may still be helpful for understanding the epidemic of COVID-19 and preventing similar infectious diseases in the future.

\section*{Conflict of interest}
The authors declare no conflict of interest.

\section*{Acknowledgment}
We acknowledged the financial supports from the National Natural Science Foundation of China (Grants No. 21877070, 11801020), the Startup Research Funding of Minjiang University (mjy19033), the Special Project of COVID-19 Epidemic Prevention and Control by Fuzhou Science and Technology Bureau (2020-XG-002), and the Fundamental Research Funding of Beijing University of Technology  (006000546318505, 006000546319509, 006000546319526).

%This section of compliance with ethics guidelines should be included separately in each article just before the reference list.

\section*{Author contributions}
L.H. designed the project. W.Y. and D.Z. collected the data. All authors analyzed the data. L.P., W.Y., C.Z. and L.H. wrote the manuscript, and all authors reviewed it.

\section*{Additional Information}
All related data and code for this work are publicly available online at https://github.com/THU-ZCAM/2019-nCoV and https://github.com/THU-ZCAM/SEIRHD-difference-model.
Applications of the current work to other countries can be found at https://github.com/ECheynet/SEIR. %https://ww2.mathworks.cn/matlabcentral/fileexchange/74545-generalized-seir-epidemic-model-fitting-and-computation.

%\section*{Appendix A. Supplementary materials}
%To rapidly share the research findings (arXiv: https://arxiv.org/abs/2002.06563), we post the data and code online as supplementary materials to this article, which can be found respectively at https://github.com/THU-ZCAM/2019-nCoV and https://github.com/THU-ZCAM/SEIRHD-difference-model.

%\bibliographystyle{unsrt}
%%abbrvnat
%%unsrt
%\bibliography{SI}
%----------------------------------------------------------------------------------------

%\end{CJK*}
\end{document}